\documentclass[preprint,12pt,number]{elsarticle} 
\usepackage{graphicx}  
\usepackage{dcolumn}   
\usepackage{bm}        
\usepackage{amssymb}   
\usepackage{xcolor}
\usepackage{natbib}
\usepackage{ulem}
\journal{Physics Letters B}

\newcommand{\be}{\begin{equation}}
\newcommand{\ee}{\end{equation}}
\newcommand{\ba}{\begin{eqnarray}}
\newcommand{\ea}{\end{eqnarray}}

\newcommand{\de}{\mbox{d}}

\begin{document}

\author[kcl]{Marco de Cesare}
\ead{marco.de\_cesare@kcl.ac.uk}

\author[kcl]{Mairi Sakellariadou}
\ead{mairi.sakellariadou@kcl.ac.uk}

\address[kcl]{Department of Physics, King's College London, University
  of London,\\Strand WC2R~2LS, London, United Kingdom}

\title{Accelerated expansion of the Universe without an inflaton and
  resolution of the initial singularity from Group Field Theory condensates}

\begin{abstract}
We study the expansion of the Universe using an effective Friedmann
equation obtained from the dynamics of GFT (Group Field Theory) isotropic condensates.
 The
  evolution equations are classical, with quantum correction terms
  to the Friedmann equation
  given in the form of effective fluids coupled to the emergent
  classical background.
  The occurrence of a bounce, which
  resolves the initial spacetime singularity, is shown to be a general
  property of the model.
   A promising feature of this model is the occurrence of an era of
accelerated expansion, without the need to introduce an inflaton
  field with an appropriately chosen potential. We discuss possible viability
issues of this scenario as an alternative to inflation.
\end{abstract}

\begin{keyword}
Quantum Gravity \sep Cosmology \sep Bounce \sep Group Field Theory
\PACS 04.60.Pp \sep 04.60.Bc \sep 98.80.-k

preprint KCL-PH-TH/2016-11  arXiv: 1603.01764
\end{keyword}

\let\today\relax

\maketitle
\section{Introduction}
Inflation, despite its undoubtful success in explaining
cosmological data and the numerous models studied in the literature,
still remains a paradigm in search of a theory.  The inflationary era
should have occured at the very early stages of our Universe, however
the inflationary dynamics are commonly studied in the context of
Einstein's classical gravity and assuming the existence of a classical
scalar field with a particularly tuned potential. Clearly, the onset
of inflation~\cite{Calzetta:1992bp,Calzetta:1992gv} and the
inflationary dynamics must be addressed within a quantum gravity
proposal. In this letter, employing results from Group Field
Theory~\cite{Oriti:2011jm,Gielen:2016dss} (GFT), we attempt to bridge the gap between
the quantum gravity era and the standard classical cosmological
model. In particular, in the context of GFT we propose a model that
can account for an early accelerated expansion of our Universe in the
absence of an inflaton field. We hence show that modifications in the
gravitational sector of the theory can account for its early stage
dynamics. Indeed, it is reasonable to expect that quantum gravity
corrections at very early times -- when geometry, space and time lose
the meaning we are familiar with -- may effectively lead to the same
dynamics as the introduction of a hypothetical inflaton field with a
suitable potential to satisfy cosmological data.

Group Field Theory is a non-perturbative and background independent
approach to quantum gravity. In GFT, the fundamental degrees of
freedom of quantum space are associated to graphs labelled by
algebraic data of group theoretic nature. The quantum spacetime is
seen as a superposition of discrete quantum spaces, each one generated
through an interaction of fundamental building blocks (called ``quanta
of geometry''), typically considered as tetrahedra. In the continuum
classical limit, one then expects to recover the standard dynamics of
General Relativity. In this sense, the notion of spacetime geometry,
gravity and time can be seen as emergent phenomena.  Group Field
Theory cosmology is built upon the existence of a condensate state of
GFT quanta, interpreted macroscopically as a homogeneous universe.

\section{GFT Cosmology}
In this work we study the properties of solutions of the modified
Friedmann equation~\cite{Oriti:2016ueo}, obtained
within the context of GFT condensates. The condensate wave
function can be written as $\sigma_j=\rho_j e^{i \theta_j}$, 
  where $j$ is a representation index. Evolution is purely relational,
thus all dynamical quantities are regarded as functions of a massless
scalar field $\phi$. Derivatives with respect to $\phi$ will be
denoted by a prime. There is a conserved charge associated to
$\theta_j$:
\be \rho_j^2\theta^{\prime}_j =Q_{j}.  \ee 
The modulus satisfies the equation of motion
\be\label{eq:EOM}
\rho_j^{\prime\prime}-\frac{Q^2}{\rho_j^3}- m_j^2 \rho_j =0, \ee 
leading to another conserved current, the \textit{GFT energy}:
\be\label{eq:GFT energy}
E_j=(\rho_j^{\prime})^2+\frac{Q_j^2}{\rho_j^2}-m_{j}^2\rho_j^2~,
\ee
where $m_j^2$ can be expressed in terms of coefficients in the
corresponding GFT theory, see Ref.~\cite{Oriti:2016ueo} for details.
Equation (\ref{eq:EOM}) admits the following solution
\be\label{eq:rho} \rho_j(\phi)=\frac{e^{(-b-\phi ) \sqrt{m_j^2}}
  \Delta(\phi)}{2 \sqrt{m_j^2}}, \ee 
where
\be
 \Delta(\phi)=\sqrt{a^2-2 a e^{2 (b+\phi ) \sqrt{m_j^2}}+e^{4 (b+\phi )
      \sqrt{m_j^2}}+4 m_j^2 Q_j^2}
\ee
and $a$, $b$ are integration constants.  From Eq.~(\ref{eq:GFT
  energy}) follows 
\be E_j=a, \ee 
whereas the charge $Q_j$ contributes to the canonical momentum of the
scalar field (see Ref.~\cite{Oriti:2016ueo})  
\be\label{eq:Qmomentum} \sum_j Q_j=\pi_{\phi}.  \ee 
The dynamics of macroscopic observables is defined through that of the
expectation values of the corresponding quantum operators. In GFT, as
in Loop Quantum Gravity, the fundamental observables are geometric
operators, such as areas and volumes. The volume of space at a given
value of relational time $\phi$, is thus obtained from the condensate
wave function as 
\be V=\sum_{j} V_{j}\rho_j^2,  \ee
 where $V_j\propto j^{3/2} \ell_{Pl}$ is the eigenvalue of the
volume operator corresponding to a given representation $j$. Using
this as a definition and differentiating w.r.t. relational time $\phi$
one obtains, as in Ref.~\cite{Oriti:2016ueo} the following equations,
which play the r\^ole of effective Friedmann (and acceleration)
equations describing the dynamics of the cosmos as it arises from that
of a condensate of spacetime quanta 
\ba \frac{V^{\prime}}{V}&=&\frac{2
  \sum_{j} V_{j}\rho_j\rho_j^{\prime}}{\sum_{j}
  V_{j}\rho_j^2},\\ \frac{V^{\prime\prime}}{V}&=&\frac{2 \sum_{j}
  V_{j}\left(E_j+2m_j^2\rho_j^2\right)}{\sum_{j} V_{j}\rho_j^2}.
\ea 
In the context of GFT, spacetime
is thus seen to emerge in the hydrodynamic limit of the theory; 
the evolution of a homogeneous
and isotropic Universe is completely determined by that of its volume.
Notice that the above equations are written in terms of functions of
$\phi$. In fact, as implied by the background independence of GFT, and
more in general of any theory of quantum geometry, \textit{a priori}
there is no spacetime at the level of the microscopic theory
and therefore no way of selecting a coordinate
time. 
Nevertheless,
we will show how it is possible to introduce
a preferred choice of time, namely proper time, in order to study the 
dynamics of the model in a way similar to the one followed for standard
homogeneous and isotropic models.
This will be particularly useful for the study of the accelerated expansion
of the Universe.
In the following we will restrict
our attention to the case in which the condensate belongs to one
particular representation of the symmetry group. This special case can
be obtained from the equations written above by considering a
condensate wave function $\sigma_j$ with support only on
$j=j_0$. Representation indices will hereafter be omitted. Hence, we
have 
\ba \frac{V^{\prime}}{V}&=&2\label{eq:effective dynamics}
\frac{\rho^{\prime}}{\rho}\equiv2g(\phi),\\ \frac{V^{\prime\prime}}{V}&=&2\label{eq:effective dynamics accel}
\left(\frac{E}{\rho^2}+2m^2\right).  \ea 
As $\phi\to\pm\infty$, $g(\phi)\to \sqrt{m^2}$ and the standard
Friedmann and acceleration equations with a constant
gravitational coupling and a fluid with a stiff equation of state are recovered. 
 We will introduce proper time by means of
the relation between velocity and momentum of the scalar field
\be\label{eq:momentum}
\pi_{\phi}=\dot{\phi}V.
\ee
Furthermore, we can \textit{define} the scale factor as the cubic root
of the volume
\be\label{eq:scaleFactor}
a\propto V^{1/3}.
\ee
We can therefore write the evolution equation of the Universe 
 obtained from GFT in the form of an \textit{effective Friedmann
  equation} ($H=\frac{\dot{V}}{3V}$ is the Hubble expansion rate and
$\varepsilon=\frac{\dot{\phi}^2}{2}$ the energy density) 
\be\label{eq:effective Friedmann}
H^2=\left(\frac{V^{\prime}}{3V}\right)^2\dot{\phi}^2=\frac{8}{9}g^2
\varepsilon.  \ee 
Using Eqs.~(\ref{eq:GFT energy}),~(\ref{eq:Qmomentum}),~(\ref{eq:momentum})
we can recast  Eq.~(\ref{eq:effective Friedmann}) in the following form
\be\label{eq:FriedmannEffFluids}
H^2=\frac{8}{9}Q^2\left(\frac{\gamma_m}{V^2}+\frac{\gamma_E}{V^3}+\frac{\gamma_Q}{V^4}\right),
\ee
where we introduced the quantities
\be\label{eq:fluids}
\gamma_m=\frac{m^2}{2},\hspace{1em}\gamma_E=\frac{V_jE}{2},\hspace{1em}\gamma_Q=-\frac{V_j^2Q^2}{2}.
\ee
The first term in Eq.~(\ref{eq:FriedmannEffFluids}) is, up to a constant factor, the energy density of a massless scalar field on a conventional FLRW background,
whereas the others represent the contribution of effective fluids with distinct equations of state and express
departures from the ordinary Friedmann dynamics. Respectively, the equations of state of the terms in Eq.~(\ref{eq:fluids})
are given by $w=1,2,3$, consistently with the (quantum corrected) Raychaudhuri equation Eq.~(\ref{eq:accelerationFluids}).
Effective fluids have been already considered in the context of LQC as a a way to encode quantum corrections, see \textit{e.g.} \cite{Singh:2005km}.

This equation  reduces to the conventional
Friedmann equation in the large $\phi$ limit, where the contributions of
the extra fluid components are negligible
\be\label{eq:Friedmann}
H^2=\frac{8\pi G}{3}\varepsilon.  \ee 
Thus, consistency in the limit demands $m^2= 3\pi G$, which
puts some constraints on the parameters of the microscopic model based
on its macroscopic limit (see Ref.~\cite{Oriti:2016ueo}).

The interpretation of our model is made clear by Eq.~(\ref{eq:FriedmannEffFluids}).
In fact the dynamics has the usual Friedmann form with a classical background represented
by the scale factor $a$ and quantum geometrical corrections
given by two effective fluids, corresponding to the two conserved quantities $Q$ and $E$.
In the following we will consider for convenience
Eq.~(\ref{eq:effective Friedmann}) in order to study the properties of solutions.

\begin{figure}
 \includegraphics[width=\columnwidth]{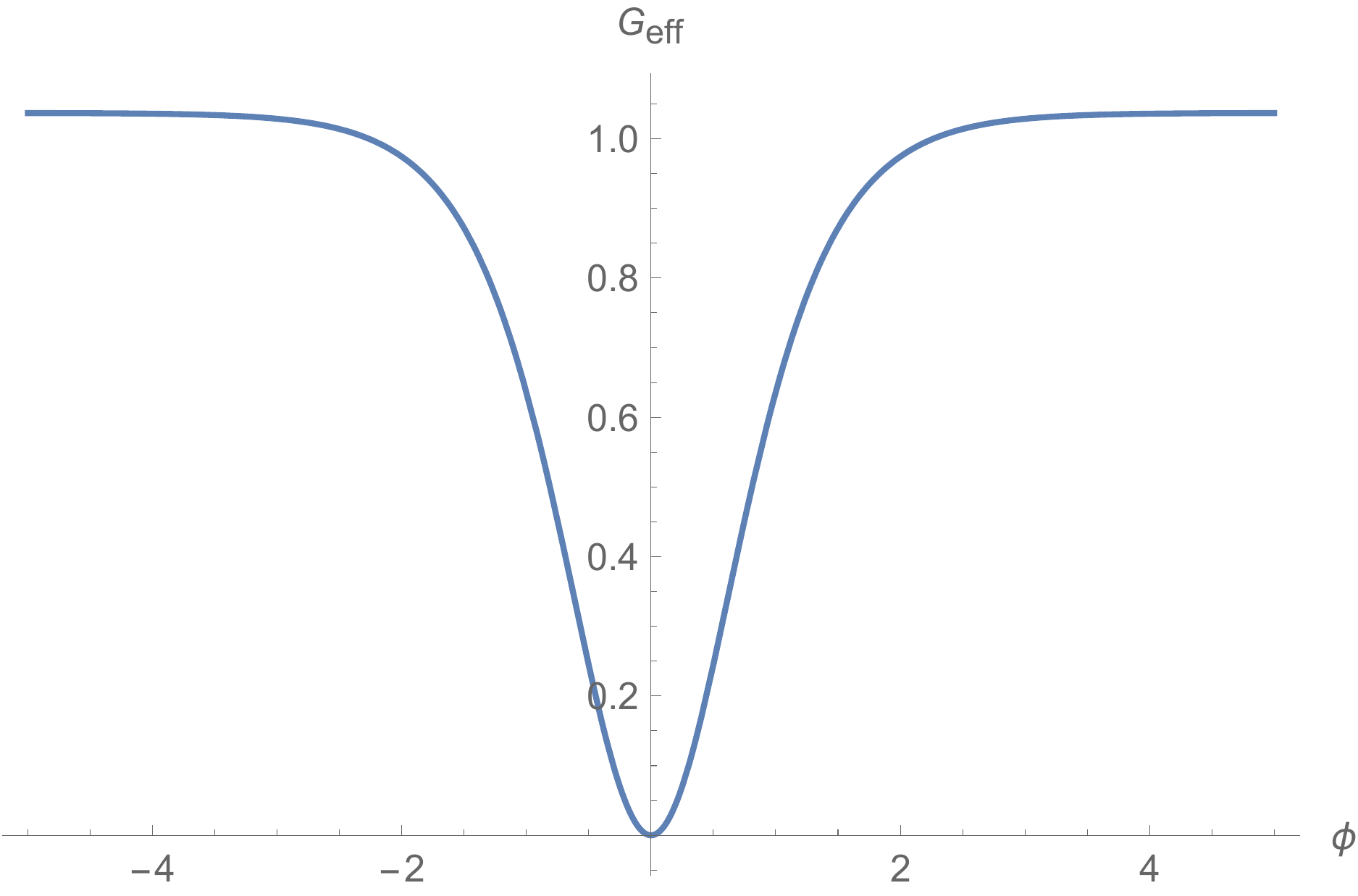}
 \caption{}\label{fig:Geff E<0}
 \includegraphics[width=\columnwidth]{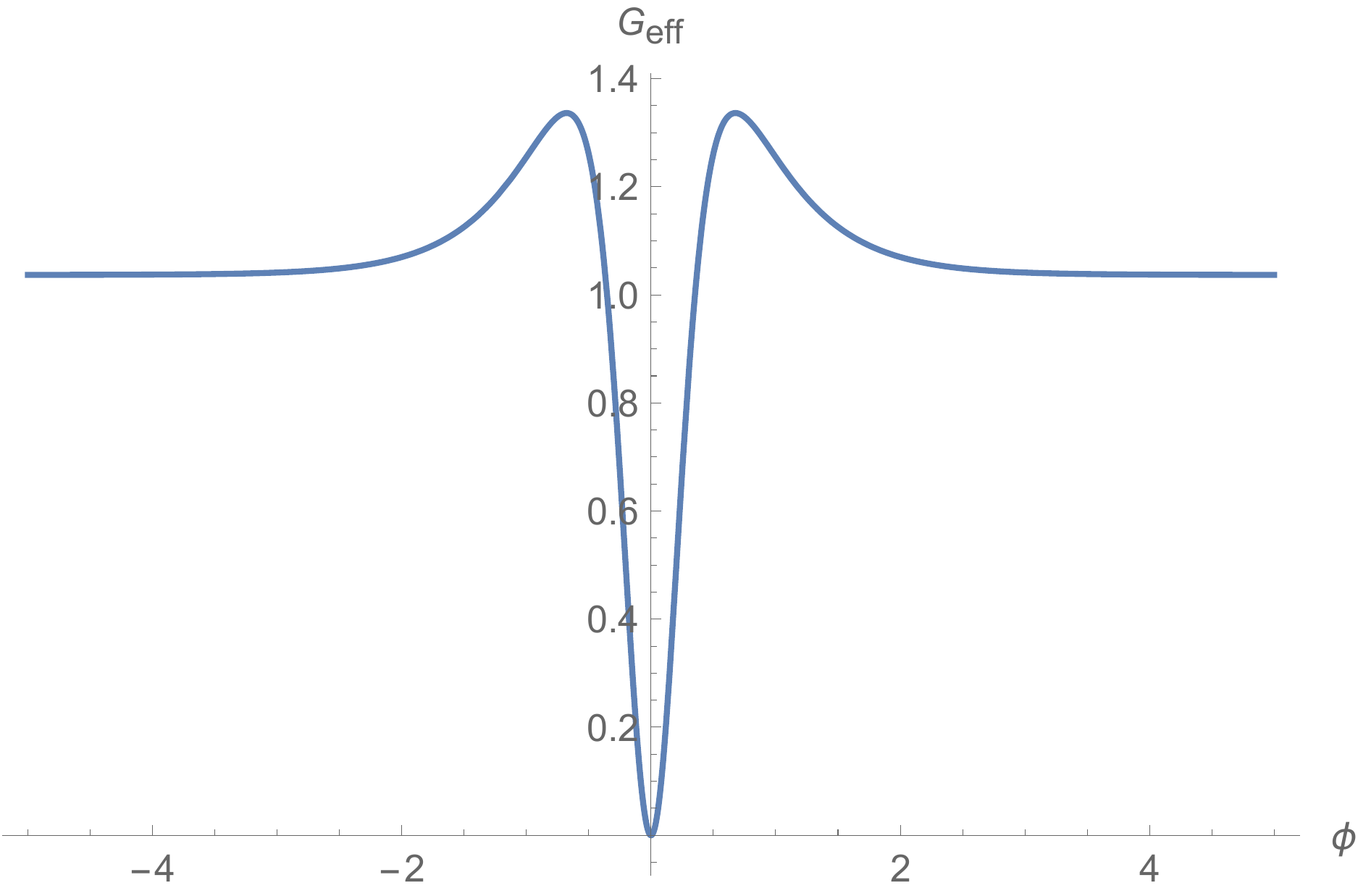}
 \caption{The effective gravitational constant as a function of
   relational time $\phi$ for $E<0$ (Fig.~\ref{fig:Geff E<0}) and $E>0$
   (Fig.~\ref{fig:Geff E>0}), in arbitrary units. There is a bounce
   replacing the classical singularity in both cases.The origin in the
   plots correponds to the bounce, occuring at $\phi=\Phi$. The
   asymptotic value for large $\phi$ is the same in both cases and
   coincides with Newton's constant. In the case $E<0$ this limit is
   also a supremum, whereas in the $E>0$ case $G_{eff}$ has two
   maxima, equally distant from the bounce, and approaches Newton's
   constant from above.}\label{fig:Geff E>0}
\end{figure}

Let us discuss in more detail the properties of the model at finite
(relational) times. Eq.~(\ref{eq:effective dynamics}) predicts a bounce
when $g(\phi)$ vanishes. We denote by $\Phi$ the ``instant'' when the
bounce takes place. One can therefore eliminate the integration
constant $b$ in favour of $\Phi$ 
\be b=\frac{\log \left(\sqrt{E^2+4
    m^2 Q^2}\right)}{2 \sqrt{m^2}}-\Phi.  \ee 
We  define the \textit{effective} gravitational constant as
\be
G_{\rm eff}=\frac{1}{3\pi}g^2, \ee
which can be expressed, using
Eqs.~(\ref{eq:rho}), (\ref{eq:effective dynamics}) as \be
G_{\rm eff}=\frac{G \left(E^2+12 \pi G Q^2\right)\sinh ^2\left(2 \sqrt{3
    \pi G} (\phi -\Phi )\right)}{\left(E -\sqrt{E^2+12 \pi G Q^2}
  \cosh \left(2 \sqrt{3 \pi G} (\phi -\Phi )\right)\right)^2}.  \ee
Its profile is given in Figs.~\ref{fig:Geff E<0},\ref{fig:Geff E>0}, in
the cases $E<0$, $E>0$ respectively. Notice that it is symmetric about
the line $\phi=\Phi$, corresponding to the bounce.

The energy density has a maximum at the bounce, where the volume
reaches its minimum value 
\be
\varepsilon_{\rm max}=\frac{1}{2}\frac{Q^2}{V_{\rm bounce}^2}, \ee 
where 
\be
V_{\rm bounce}=\frac{V_{j_0} \left(\sqrt{E^2+12 \pi G Q^2}-E\right)}{6 \pi
  G}.  \ee 
  Clearly, the singularity is always avoided for $E<0$ and,
provided $Q\neq 0$, it is also avoided in the case
$E>0$. Moreover, if the GFT energy is negative, the energy density has
a vanishing limit at the bounce for vanishing $Q$:
 \be \lim_{Q\to
  0}\varepsilon_{\rm max}=0, \hspace{1em} E<0.  \ee 
Therefore in this limiting case the energy density is zero at all
times. Nevertheless, the Universe will still expand following the
evolution equations~(\ref{eq:effective dynamics}) and 
\be \lim_{Q\to0}V(\phi)=\frac{|E|
  V_{j_0} \cosh ^2\left(\sqrt{3 \pi G } (\phi -\Phi )\right)}{3 \pi
  G}, \hspace{1em} E<0.  \ee 
This is to be contrasted with classical
cosmology~(\ref{eq:Friedmann}), where the rate of expansion is zero
when the energy density vanishes.

\begin{figure}
 \includegraphics[width=\columnwidth]{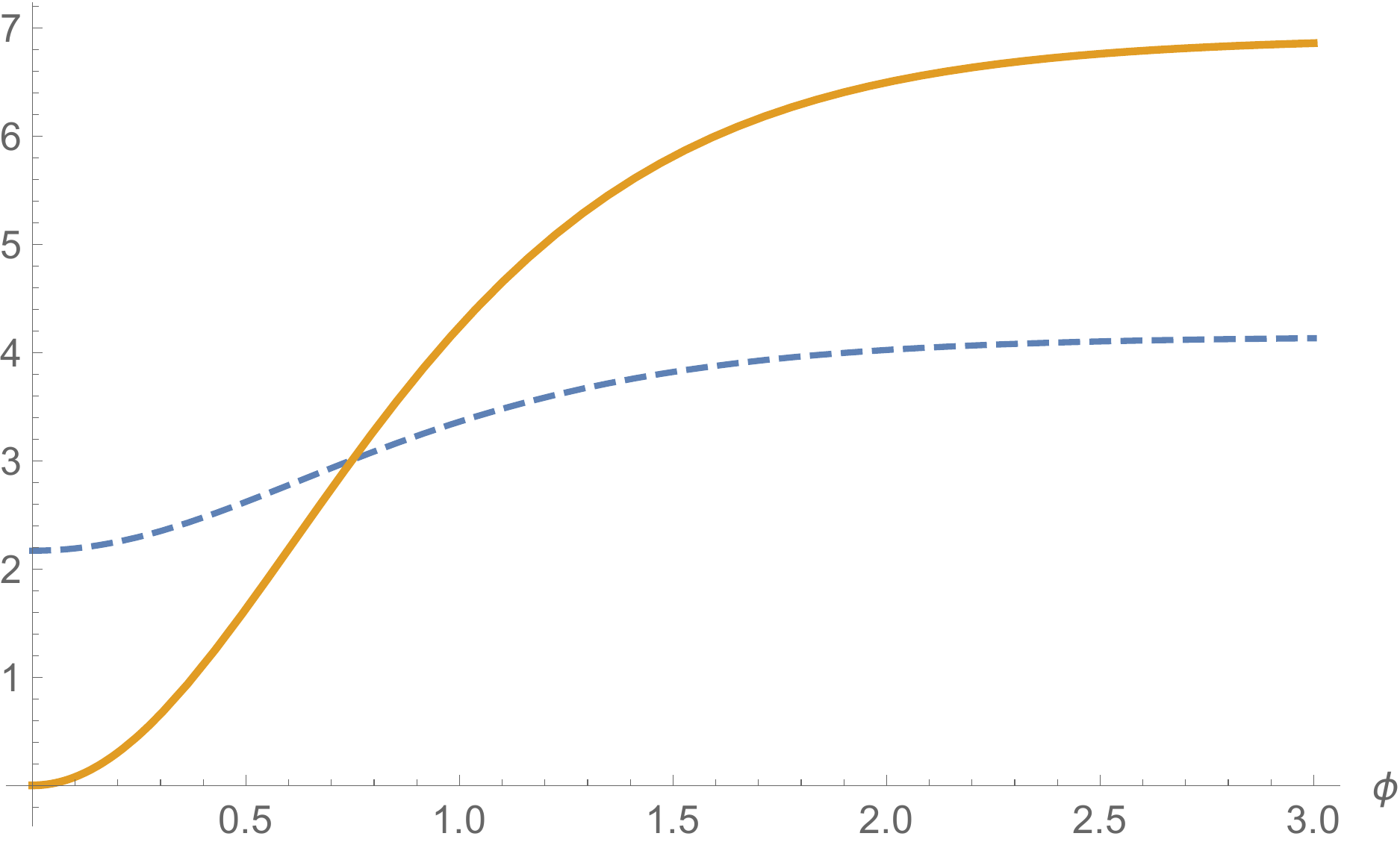}
 \caption{}\label{fig:E<0} 
 \includegraphics[width=\columnwidth]{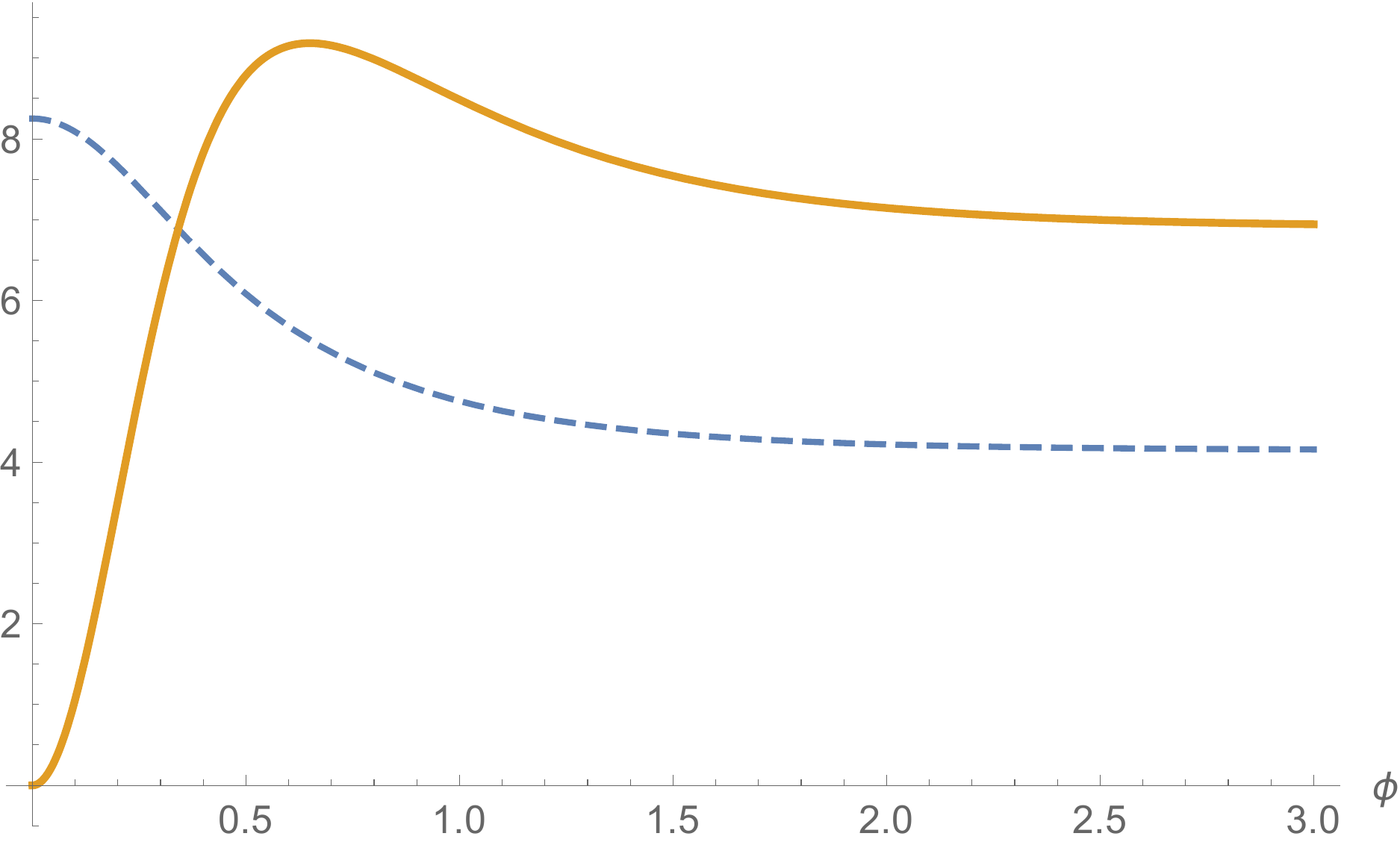}
 \caption{The l.h.s. and the r.h.s of inequality
   (\ref{ineq:acceleration}) as functions of the relational time
   $\phi$ correspond to the the dashed (blue) and thick (orange) curve
   respectively, in arbitrary units. When the dashed curve is above
   the thick one the Universe is undergoing an epoch of accelerated
   expansion following the bounce. Figure~\ref{fig:E<0} corresponds to
   the case $E<0$, whereas Fig.~\ref{fig:E>0} is relative to the
   opposite case $E>0$. Notice that for the latter there is a stage of
   maximal deceleration after exiting the ``inflationary'' era. After
   that the acceleration takes less negative values until it relaxes
   to its asymptotic value. For $E<0$ instead the asymptote is
   approached from below.}\label{fig:E>0}
 \end{figure}

It is possible to express the condition that the Universe has a
positive acceleration in purely relational terms. In fact this very
notion relies on the
choice of a particular time parameter, namely proper time, for its definition.
Introducing the scale factor
and proper time as in Eqs.~(\ref{eq:momentum}),~(\ref{eq:scaleFactor})
one finds
\be\label{eq:acceleration}
\frac{\ddot{a}}{a}=\frac{2}{3}\varepsilon\left[
  \frac{V^{\prime\prime}}{V}-\frac{5}{3}\left(\frac{V^{\prime}}{V}\right)^2\right].
\ee 
We observe that the last equation can also be rewritten as
\be\label{eq:accelerationFluids}
\frac{\ddot{a}}{a}=-\frac{4}{9}Q^2\left(4\frac{\gamma_m}{V^2}+7\frac{\gamma_E}{V^3}+10\frac{\gamma_Q}{V^4}\right).
\ee
We can trade the condition $\ddot{a}>0$
for having an accelerated expansion with the following one, which only
makes reference to relational evolution of observables.  
\be
\frac{V^{\prime\prime}}{V}>\frac{5}{3}\left(\frac{V^{\prime}}{V}\right)^2
\ee 
The two conditions are obviously equivalent. 
However,
the second one has a wider range of applicability, since it is
physically meaningful also when the scalar field has
vanishing momentum. Making use of
  Eq.~(\ref{eq:effective dynamics}) the condition above can be
rewritten as 
\be\label{ineq:acceleration}
4m^2+\frac{2E}{\rho^2}>\frac{20}{3} g^2.  \ee 
This is satisfied
trivially in a neighbourhood of the bounce since $g$ vanishes there
and the l.h.s. of the inequality is strictly positive, see
Figs.~\ref{fig:E<0},~\ref{fig:E>0}. It is instead violated at infinity,
consistently with a decelerating Universe in the classical regime.
 
 \section{Discussion}\label{sec:discussion}
 The dynamics of the Universe predicted by the GFT model is purely relational,
 \textit{i.e.}, using the language of Ref.~\cite{Rovelli:2001bz}, it is expressed by the functional relation
 between \textit{partial observables}, here given by the volume $V$ and the
 scalar field $\phi$. According to this interpretation, physically meaningful statements
 about the predicted value of $V$ can only be made in conjunction with
 statements about the predicted outcome of a measurement of $\phi$.
 In fact, this interpretation is inspired by one of the main insights of GR,
 namely by the observation that coordinate time is purely gauge-dependent, and is therefore deprived
 of any physical meaning. Thus, it cannot be expected to play any r{\^o}le
 in the quantum theory either.
 In other words, the dynamics is entirely given by the so called \textit{complete observables},
 which in this model  are exhausted by the functional relation $V(\phi)$. In a theory with gauge invariance, such quantities are the only ones having physical meaning;
 they can be seen as functions on the space of solutions modulo all gauges~\cite{Rovelli:2001bz}.
 Gauge invariance of $V(\phi)$ is trivially verified in classical cosmology;
 it is also valid at the quantum level, since gauge invariance of the volume operator
 follows from its general definition in GFT~\cite{Gielen:2016dss}. A discussion on the implementation of
  diffeomorphism invariance in GFT can be found in Ref.~\cite{Baratin:2011tg}.

The reader might be interested in finding a closer correspondence between our discussion
of relational dynamics and examples considered, \textit{e.g.}, in Ref.~\cite{rovelli1991time}. In that work
relational dynamics was obtained, both at the classical and the quantum level,
adopting a canonical formulation and constructing \textit{complete observables}
in the case of simple models. However, we must point out that the definition of \textit{partial} and \textit{complete observables}
is much more general and does not rely on a phase space structure, 
but only on the possibility of identifying gauge equivalence classes in the space of solutions\footnote{In the
canonical formalism such space is defined as the space of gauge orbits generated by the first
class constraints on the constraint surface~\cite{Rovelli:2001bz}.}. 
Furthermore, since the quantum theory
is not based on canonical quantization, it is not clear how a presymplectic structure might emerge
in the classical limit from the full theory. It should nevertheless be possible to find such a geometric
structure and a Hamiltonian at least for the cosmological sector of the theory considered here;
an investigation which however lies beyond the scope of the present work.

Our classical model is free of gauge redundancies since
Eqs.~(\ref{eq:effective dynamics}),~(\ref{eq:effective dynamics accel}) are equations of motion
for the expectation value of a gauge invariant operator in the quantum theory. Therefore, 
the complete observable $V(\phi)$ can be found by solving the equations of motion. Gauge redundancies can
nevertheless be reintroduced at the macroscopic level by means of a time parameter,
in order to make contact with the corresponding symmetry of classical cosmology. More specifically,
one could write Eq.~(\ref{eq:momentum}) with the velocity of the field
evaluated w.r.t. a time $t^{\prime}$ distinct from proper time $t$ as
\be
\pi_{\phi}=N^{-1}V\frac{\de\phi}{\de t^{\prime}}, \hspace{1em} N=\frac{\de t}{\de t^{\prime}},
\ee
using which it is possible to write the dynamical equations for all time parametrizations.
From the above it is clear how Eq.~(\ref{eq:effective Friedmann}) follows as a consequence
of the choice of a specific time parameter, or equivalently lapse function $N=1$.

We have shown that the bounce is accompanied by an early stage of accelerated expansion,
occurring for any values of the conserved quantities $E$, $Q$ (provided that the latter is non-vanishing)
 and despite the fact that no potential has
been introduced for the scalar field. This is a promising feature of the model which  indicates that the
framework adopted allows for a mechanism leading to an accelerated expansion through quantum geometry effects.
From the point of view of Eq.~(\ref{eq:effective Friedmann}) one can say that the accelerated expansion 
is a consequence of $G_{eff}$ not being constant. By looking at Eqs.~(\ref{eq:FriedmannEffFluids}),~(\ref{eq:fluids}),~(\ref{eq:accelerationFluids})
one sees that this phenomenon can be traced back to an effective fluid component having a negative energy density
arising from quantum geometry effects.
The model we considered is a very simple one and represents a first step towards a new understanding of cosmology in the GFT framework. However, there are
some caveats. In fact, a full viability of the scenario of \textit{geometric inflation}
can only be proven by showing the robustness of the result when considering more complicated GFT models
with different matter fields coupled to gravity. Only then one would be able to give a definite answer as whether the era of accelerated expansion lasts long enough to cure the shortcomings of the standard Hot Big Bang model, while it eventually leads to a radiation-dominated era though a graceful exit scenario. 
Nevertheless, despite the simplicity of the model, we expect it to provide a good description of the dynamics of the Universe
at least at the onset of inflation, where the energy density of the scalar field is supposed to dominate over all other forms of energy.

 \section{Outlook and conclusions}
 We studied the properties of a model of quantum cosmology obtained in
 Ref.~\cite{Oriti:2016ueo} in the hydrodynamic limit of GFT.   We have shown that this
 model displays significant departures from the dynamics of a classical
 FLRW spacetime. The emergent classical background satisfies an evolution equation
 of the Friedmann type with quantum corrections appearing in the r.h.s as
 effective fluids with distinct equations of state. Such correction terms
 vanish in the limit of infinite volume, where the standard Friedmann dynamics
 is recovered.
 
 The main results of this work are two. First, confirming the result of Ref.~\cite{Oriti:2016ueo},
 we have shown that there
 is a \textit{bounce}, taking place regardless of the particular values
 of the conserved charges $Q$ and $E$. It should be pointed out that
 the origin of this bounce is quite different from the one given by
 Loop Quantum Cosmology (see Refs.~\cite{Ashtekar:2006wn,Ashtekar:2007em}),
which is an independent approach based on a symmetry
 reduced quantization.

The second result is the occurence of an era of \textit{accelerated
  expansion} without the need for introducing \textit{ad hoc} potentials
and initial conditions for a scalar field. We suggest that the picture given could
replace the inflationary scenario. Since it is an inherently
quantum description of cosmology, it does not share the unsatisfactory
features of inflationary models, which were spelled out in the
introduction. However, the viability of our model
as an alternative to inflation is at this stage still an hypothesis,
which will be investigated further in future work. In fact the model
must be extended to include also other forms of energy
and to ensure that common problems of inflationary models
(as those spelled out in the last part of Section~\ref{sec:discussion}) are solved.

 We have seen that the interesting features of the model arise
 from quantum geometry corrections which are captured by
 a description in terms of effective fluids defined on the emergent classical
 background. A similar phenomenon was already observed in LQC (see \textit{e.g.} \cite{Singh:2005km}).
 In light of our results, it will be interesting to understand whether
 it is possible to relate the origin of such effective fluids
 coming from quantum geometry in LQC and GFT.
  We also showed that there is another
 way of formulating the dynamics, which
 makes no reference to such effective fluids,
 but instead differs from the standard Friedmann
 equation in that the gravitational constant is
 replaced by a dynamical quantity.
  In fact, another interesting result is that, even though Newton's constant
 is related to, and actually constrains, the parameters of the
 microscopic GFT theory (as shown in \cite{Oriti:2016ueo}), the dynamics of the
 expansion of the Universe is actually determined by the
 \textit{effective gravitational constant} $G_{\rm eff}$. 
 We should stress that such
  quantity was introduced in first place for the only purpose of
  studying the properties of solutions of the model. Nevertheless, it
  is tempting to go one step further and consider it as an effective macroscopic quantity
determined by the collective behaviour of spacetime quanta.
However, such an interpretation would possibly pose more puzzles
than it solves since, as shown in Refs.~\cite{stochasticG},~\cite{Fritzsch:2015lua}, a dynamical
gravitational constant must bear with it extra sources of energy-momentum
in order to ensure compatibility with the Bianchi identities.
Violation of the Bianchi identities would in fact imply that the
structure of the emergent spacetime is non-Riemannian. The way in which inflation
could be understood in that case is not clear and would deserve
further study.
Nevertheless, the formally equivalent description of the dynamics (as far as the Friedmann equation is concerned)
in terms of an effective gravitational constant  deserves further investigation.
 More precisely, if interpretational issues in the non-Riemannian framework can be properly addressed,
 such studies might
 shed some light on the nature of the gravitational constant and
 point out whether it actually deserves the status of fundamental
 constant, along with the possibility of measuring its time variation.

 \section*{Acknowledgement}
 The authors would like to thank Martin Bojowald for correspondence.
 MdC would also like to thank Andreas Pithis for useful discussions and
 comments on the manuscript.

\bibliographystyle{elsarticle-num.bst}
\bibliography{G}

\end{document}